\documentclass[a4paper]{spie}  
\usepackage[]{graphicx}
\usepackage{amssymb,amsmath,url,psfrag}
\usepackage{verbatim}
\usepackage{bm}

\newcommand{\Bolivarallee}{Boliva\hspace{-0.1mm}r\hspace{0.15mm}a\hspace{-0.1mm}llee}

\newcommand{\Takustrasse}{Taku\hspace{0.25mm}s\hspace{-0.1mm}tra{\ss}e}
\newcommand{\Kaiserstrasse}{Kaiser\hspace{0.25mm}s\hspace{-0.1mm}tra{\ss}e}

\title{Global optimization of complex optical structures using Baysian optimization based on Gaussian processes}

\author{
Philipp-Immanuel~Schneider,\supit{\,a}
Xavier~Garcia~Santiago,\supit{\,ab}
Carsten~Rockstuhl,\supit{\,b}
Sven~Burger\supit{\,ac}
\skiplinehalf
\supit{a}
JCMwave GmbH,
\Bolivarallee~22, 
D\,--\,14\,050 Berlin,
Germany
\smallskip\\
\supit{b}
Karlsruher Institut f\"ur Technologie\,(KIT),
\Kaiserstrasse~7,
D\,--\,76\,131 Karlsruhe,
Germany
\smallskip\\
\supit{c}
Zuse Institute Berlin\,(ZIB),
\Takustrasse~7,
D\,--\,14\,195 Berlin,
Germany
\authorinfo{
Corresponding author: P.-I.~Schneider\\
URL: http://www.jcmwave.com\\
URL: http://www.zib.de
}}

\begin{document}
\maketitle
\noindent
This paper will be published in Proc.~SPIE Vol.~{\bf 10335}
(2017) 103350O ({\it Digital Optical Technologies}, DOI: 10.1117/12.2270609)
and is made available 
as an electronic preprint with permission of SPIE. 
One print or electronic copy may be made for personal use only. 
Systematic or multiple reproduction, distribution to multiple 
locations via electronic or other means, duplication of any 
material in this paper for a fee or for commercial purposes, 
or modification of the content of the paper are prohibited.
Please see original paper for images at higher resolution. 

\begin{abstract}
  Numerical simulation of complex optical structures enables their  optimization with respect to specific objectives. Often, optimization is done by multiple successive parameter scans, which are time consuming and computationally expensive. We employ here Bayesian optimization with Gaussian processes in order to automatize and speed up the optimization process.
  As a toy example, we demonstrate optimization of the shape of a free-form reflective meta surface such that it diffracts light into a specific diffraction order.
  For this example, we compare the performance of six different Bayesian optimization approaches with various acquisition functions and various kernels of the Gaussian process. 
\end{abstract}


\section{Introduction}

Modern technologies such as laser writing~\cite{LPOR:LPOR201100046} or electron-beam lithography~\cite{Chen201557} allow for the manufacturing of micro and nano optical structures with an increasing degree of accuracy and flexibility. In order to create structures that are optimized for specific purposes, numerical solutions of Maxwell's equations are an important tool. The optimal values of the design parameters are often identified by parameter scans, i.e., by performing numerical simulations for equidistant values of the design parameters. In practice, constant-step parameter scans are only applicable for one or two parameters at a time since a three-dimensional parameter scan with, say, 20 sampling points in each direction already requires $20^3 = 8000$ simulations. Hence, one often optimizes optical structures with many degrees of freedom by performing consecutive single-parameter scans. This is a time consuming approach, which, in general, does not converge to the global optimum.

In order to improve the design process, we apply Bayesian optimization (BO) techniques that are based on Gaussian processes. BO is a strategy for the optimization of black-box objective functions that are hard to evaluate. It is applied in a wide range of design problems including robotics, combinatorial optimisation, experimental design, or environment monitoring~\cite{shahriari2016taking}. BO is based on a stochastic model of the objective function. The model is exploited to decide where to evaluate the function next with the ultimate goal to reach towards a global minimum (or maximum). The information from each evaluation is included in the stochastic model. In contrast to local optimization strategies such as gradient descent or Newton algorithms BO uses information from all previous evaluations to make better search decisions. While this comes at the cost of more complex computations for making these decisions, it enables a global search strategy. Moreover, regarding rigorous simulations of complex optical structures, the evaluation of the objective function takes usually much longer than the computation of any search decision such that BO does not slow down the search significantly. 

We demonstrate in this contribution the usage of BO techniques by optimizing the geometry of a periodic reflective meta surface with  smooth sub-wavelength structuring. The design objective is to diffract as much light as possible  into the first diffraction order.

This paper is structured as follows: In Section~\ref{sec:BO} we give a short introduction on Bayesian optimization and Gaussian processes. A more detailed introduction on BO is given, e.g., by Sha\-hari~\emph{et al.}~\cite{shahriari2016taking}. For an overview of Gaussian processes, see Rasmussen and Williams~\cite{williams2006gaussian}. In Section~\ref{sec:Surface} we introduce the objective function and the parametrization of the test geometry. We employ different BO strategies for finding optimal geometry parameters in Section~\ref{sec:experiments} and conclude in Section~\ref{sec:conclusion}.

\section{Bayesian Optimization with Gaussian Processes}
\label{sec:BO}

The goal of Bayesian optimization (BO) is to identify the minimum (or maximum) of an unknown objective function $f$:
\begin{equation}
\mathbf{x^*} = \underset{\mathbf{x} \in \mathcal{X}}{\operatorname{arg\,min}} f(\mathbf{x}),
\end{equation}
where $\mathcal{X}$ is the design space. In the following, we consider the case of an unperturbed observation of $f$ for every $\mathbf{x}\in\mathcal{X}$ where $\mathcal{X} \subset \mathbb{R}^d$ is a $d$-dimensional hyper-cube. In general, BO can be applied to spaces that involve categorical parameters or constrained parameter spaces and it can deal with function obervations that are perturbed by stochastic noise~\cite{shahriari2016taking}.

The idea of BO is to build a stochastic model of the objective function $f$ that can be updated with information from previous evaluations of $f$ and that is used to drive the optimization strategy. In the following, we will employ Gaussian processes due to their flexibility and tractability. The optimization strategy itself is determined by an \emph{acquisition function}, which is used to determine the next point to evaluate.

\begin{figure}[htp]
\centering
\includegraphics[width=0.47\linewidth]{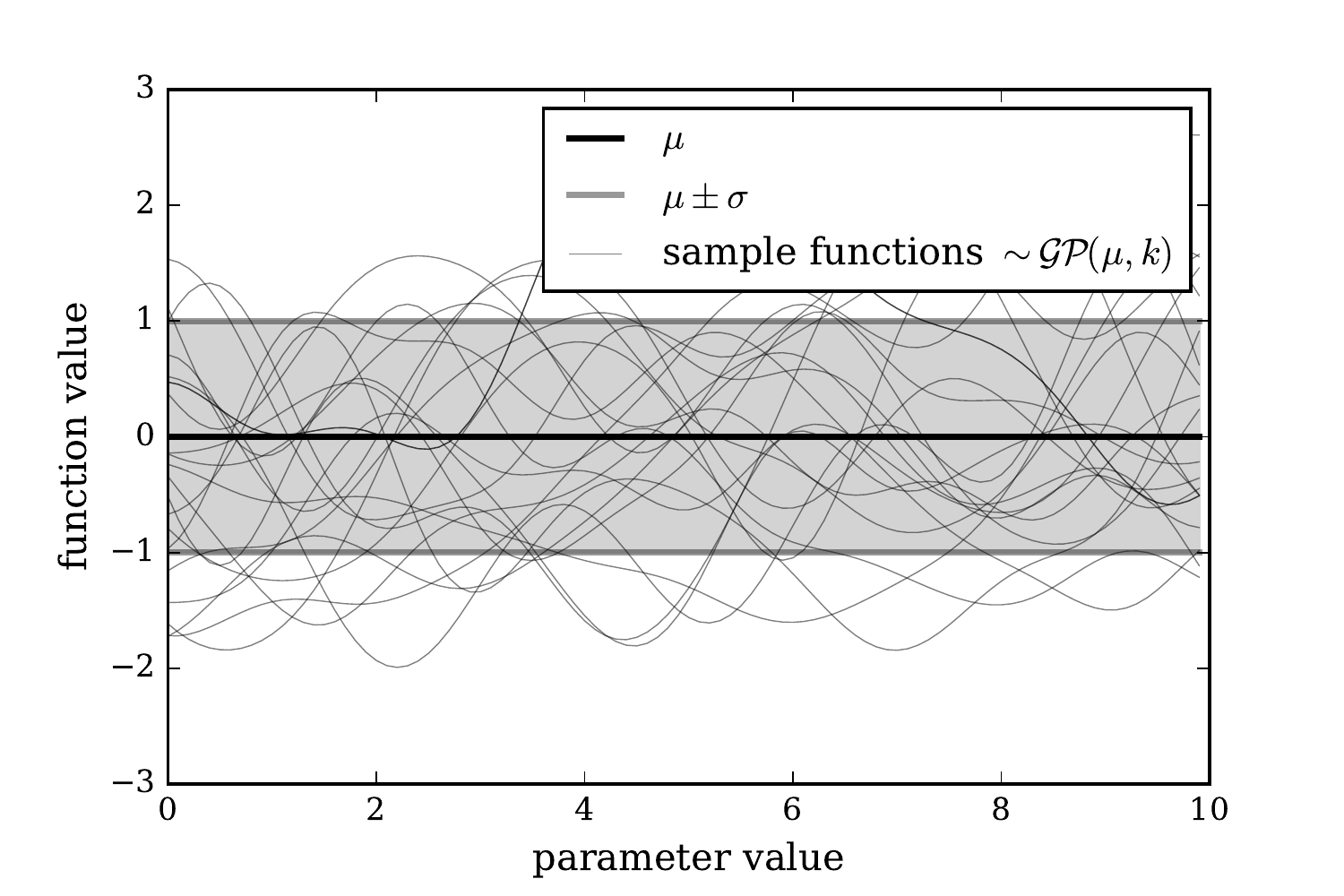}
\includegraphics[width=0.47\linewidth]{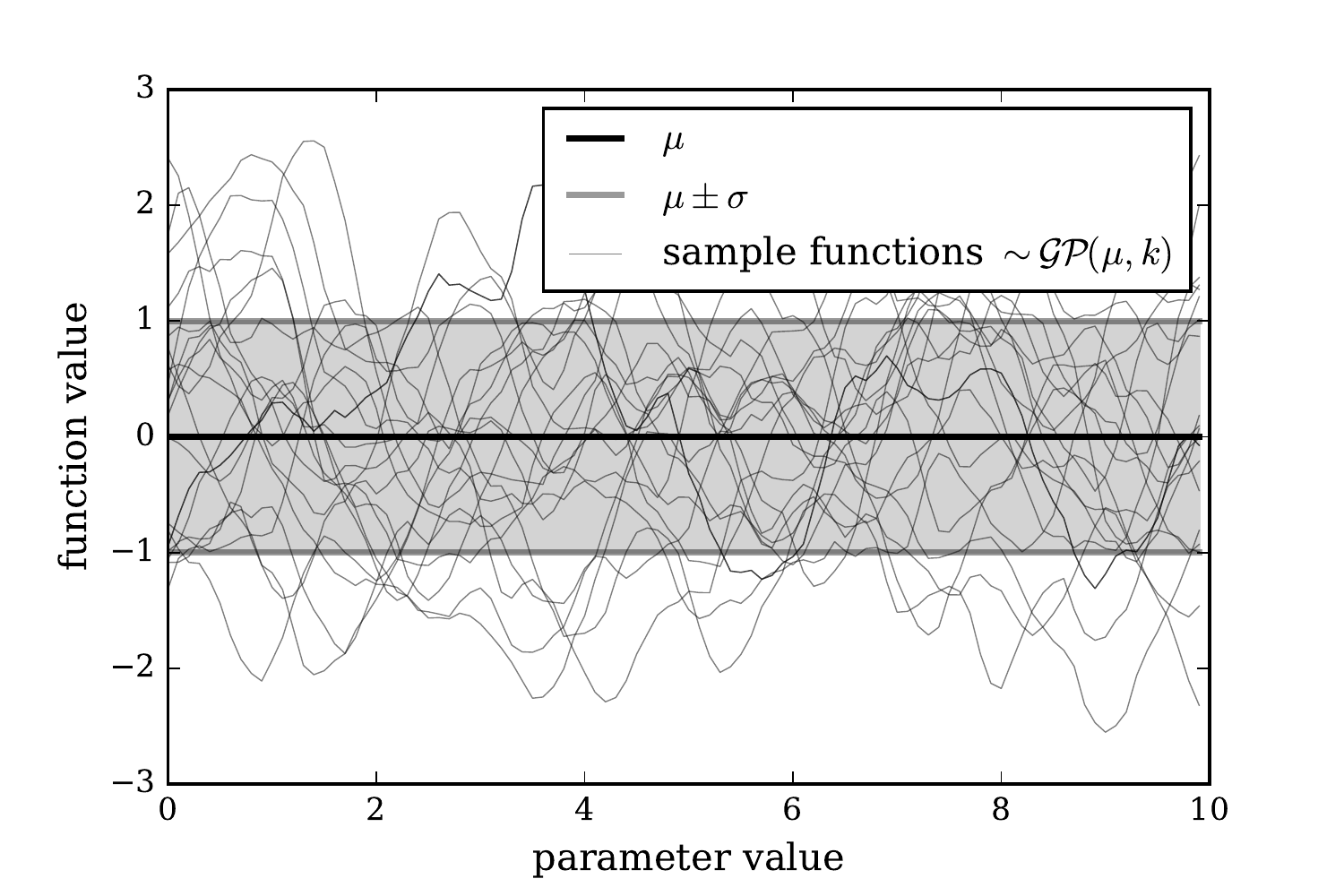}
\caption{\textbf{Left:} Random samples drawn from a Gaussian process (see Eq.~\eqref{eq:PY}) with a zero mean function ($\mu(x) = 0$) and a squared exponential kernel $k = k_{\rm SE}$ with hyper-parameters $\sigma = l$ and $l_1 = 1$. \textbf{Right:} The same with a Mat\'{e}rn $5/2$ kernel $k = k_{\rm M52}$. The squared exponential kernel results in much smoother functions than the Mat\'{e}rn $5/2$ kernel.}
\label{fig:Kernels}
\end{figure}

\subsection{Gaussian process}

A Gaussian process (GP) is a statistical model where observations occur in a continuous domain, e.g., time or space. Every point in the input space is associated with a normally distributed random variable. A statistical distribution $G_\mathbf{x}$ is a Gaussian process (notation: $G_\mathbf{x} \sim \mathcal{GP}(\mu,k)$) if for any $N$ points $\mathbf{x}_1,\cdots \mathbf{x}_N \in \mathcal X$ the tuple $\mathbf{Y}=(G_{\mathbf{x}_1}, \cdots, G_{\mathbf{x}_N})$ is a multivariate Gaussian random variable that is completely defined by a mean function $\mu : \mathcal{X} \rightarrow \mathbb{R}$ and a positive definite covariance function $k : \mathcal{X}\times \mathcal{X} \rightarrow \mathbb R$. The probability distribution of the tuple $\mathbf{Y}$ is given as
\begin{equation}
\label{eq:PY}
 P(\mathbf{Y}) = \frac{1}{(2\pi)^{N/2}|\mathbf{\Sigma}|^{1/2}}\exp\left[-\frac{1}{2}(\mathbf{Y}-\pmb{\mu})^T\mathbf{\Sigma}^{-1}(\mathbf{Y}-\pmb{\mu})\right] 
\end{equation}
(notation: $\mathbf{Y} \sim \mathcal{N}(\pmb{\mu},\mathbf{\Sigma})$). The mean function determines the vector of mean values $\pmb{\mu} = \left[\mu(\mathbf{x}_1),\cdots,\mu(\mathbf{x}_k)\right]^T$ and the covariance function determines the covariance matrix $\mathbf{\Sigma} = \left[k(x_i,x_j)\right]_{i,j}$.

By choosing different kernels $k(\mathbf{x},\mathbf{x}')$ a large class of random functions can be described. In the following we consider two different kernels, which are widely used in practice: the square exponential kernel
\begin{equation}
\label{eq:SE}
k_{\rm SE}(\mathbf{x},\mathbf{x}') = \sigma^2 \exp\left(\frac12 r^2(\mathbf{x},\mathbf{x}')\right)\;\;\text{with}\;\; 
r^2(\mathbf{x},\mathbf{x}') = \sum_{i=1}^d (x_i - x_i')^2/\l_d^2
\end{equation}
and the Mat\'{e}rn $5/2$ kernel
\begin{equation}
\label{eq:M52}
k_{\rm M52}(\mathbf{x},\mathbf{x}') = \sigma^2\left(1 + \sqrt{5 r^2(\mathbf{x},\mathbf{x}')} + \frac{5}{3}r^2(\mathbf{x},\mathbf{x}') \right)  \exp\left(-\sqrt{5 r^2(\mathbf{x},\mathbf{x}')}\right).
\end{equation}
The hyper-parameters $\sigma$ and $l_1,l_2,\cdots$ determine the standard deviation and the length scales of the random distributions.
Figure~\ref{fig:Kernels} shows multiple random samples drawn from Eq.~\eqref{eq:PY} for densely chosen input parameters $x_1,x_2,\cdots \in [-10,10]$. As one can see, the squared exponential kernel (left image) results in much smoother curves than the Mat\'{e}rn $5/2$ kernel (right image). Hence, a GP with a squared exponential kernel is suitable for modeling very smooth objective functions while a Mat\'{e}rn $5/2$ models objective functions with more abrupt value changes.

\begin{figure}[htp]
\centering
\includegraphics[width=0.47\linewidth]{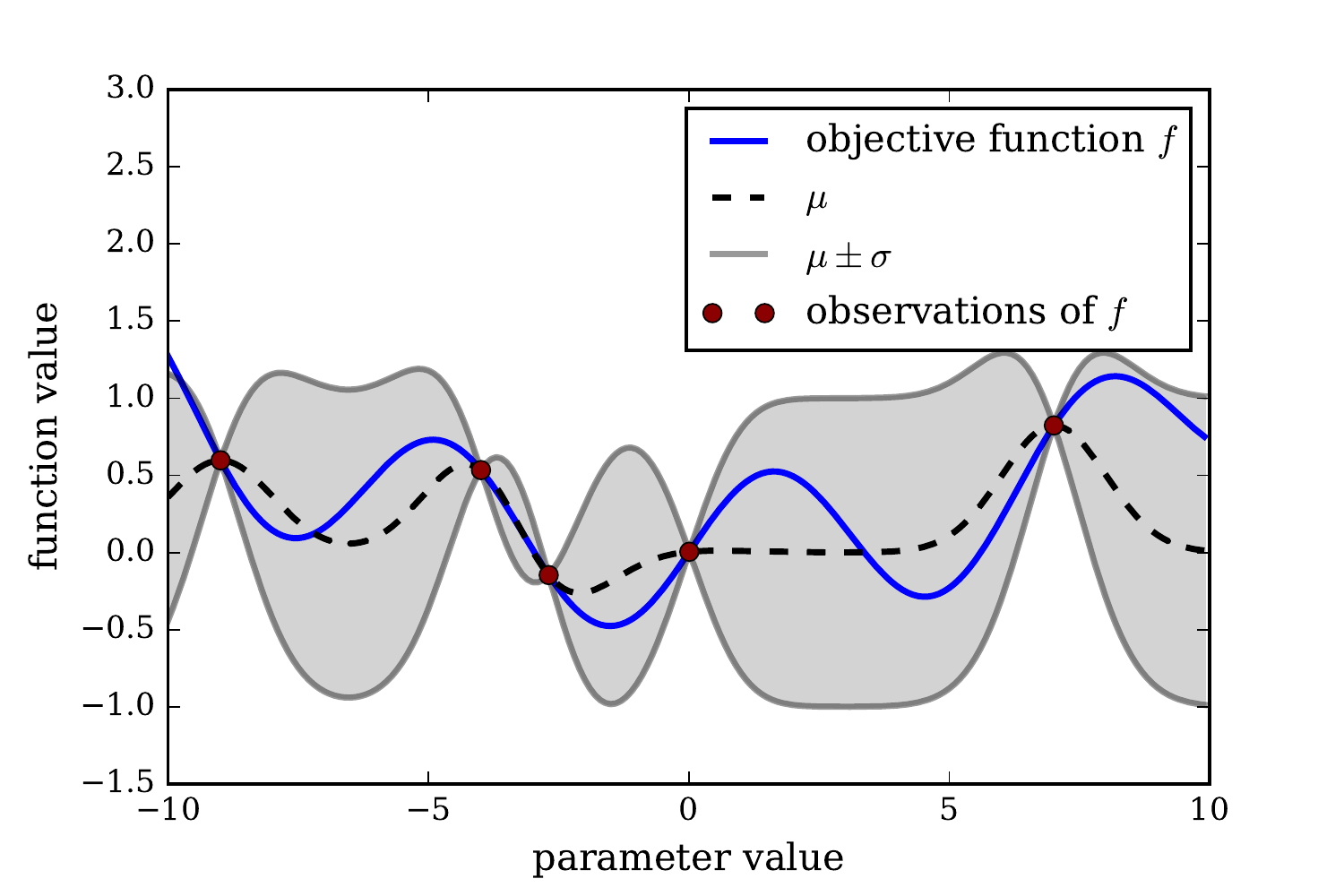}
\includegraphics[width=0.47\linewidth]{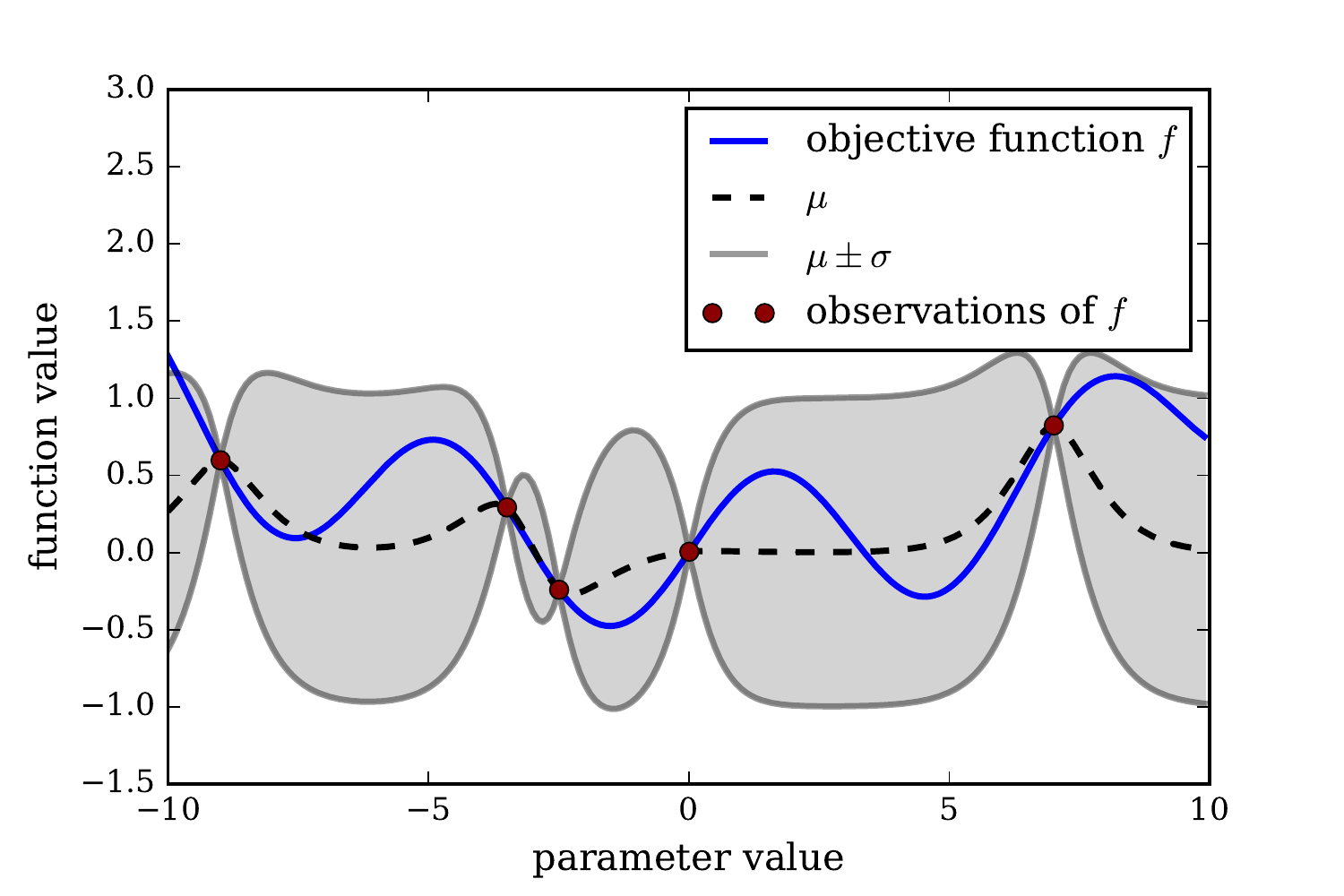}
\caption{\textbf{Left:} Mean and standard deviation of the Gaussian process with a squared exponential kernel defined in Eq.~\eqref{eq:SE} after five evaluations of the sample objective function $f(x) = 0.5 \sin(x) + 0.01 x^2$. 
\textbf{Right:} The same with the Mat\'{e}rn $5/2$ kernel defined in Eq.~\eqref{eq:M52}.
The hyper-parameters of the kernels are the same as in Fig.~\ref{fig:Kernels}.
The two Gaussian processes differ only slightly. For example, the squared exponential kernel leads to a slightly smoother behavior of the mean function.
}
\label{fig:Regression}
\end{figure}

\subsection{Gaussian process regression}

Suppose $M$ values $\mathbf{Y} = (y_1,\cdots y_N) = (f(x_1),\cdots,f(x_M))$ of the objective function are known. Then the posterior distribution will also be a Gaussian process $G'$. Again, any tuple $\mathbf{Y}^*=(G'_{\mathbf{x}^*_1}, \cdots, G'_{\mathbf{x}^*_N})$ is a multivariate Gaussian random variable with the probability distribution
\begin{equation}
\label{Eq:regression}
\mathbf{Y}^*|\mathbf{Y} \sim \mathcal{N}(\pmb{\mu}_1 + \mathbf{\Sigma}_{12}\mathbf{\Sigma}_{22}^{-1}(\mathbf{Y}-\pmb{\mu}_2), \mathbf{\Sigma}_{11} - \mathbf{\Sigma}_{12}\mathbf{\Sigma}^{-1}_{22}\mathbf{\Sigma}_{21})
\end{equation}
with $\pmb{\mu}_1 = [\mu(\mathbf{x}_1^*),\cdots,\mu(\mathbf{x}_N^*)]^T$, $\pmb{\mu}_2 = [\mu(\mathbf{x}_1),\cdots,\mu(\mathbf{x}_M)]^T$, $(\mathbf{\Sigma}_{11})_{i,j} = k(\mathbf{x}_i^*,\mathbf{x}_j^*)$, $(\mathbf{\Sigma}_{12})_{i,j} = k(\mathbf{x}_i^*,\mathbf{x}_j)$, $\mathbf{\Sigma}_{21} = {\mathbf{\Sigma}_{12}}^T$, and $(\mathbf{\Sigma}_{22})_{i,j} = k(\mathbf{x}_i,\mathbf{x}_j)$. 

Fig.~\ref{fig:Regression} shows a Gaussian process regressions for five known function values with the kernel functions $k_{\rm SE}$ and $k_{\rm M52}$, respectively.

\subsection{Optimization of hyper-parameters}
Initially, the hyper-parameters of the GP $\omega = (\sigma,l_1,l_2,\cdots)$ are unknown and are usually initialized with some reasonable values\footnote{An alternative approach is to treat the hyper-parameters themselves as random variables with some prior distribution and marginalize every optimization strategy over the hyper-parameters.\cite{snoek2012practical}}. 
Having drawn some samples of the objective function one can optimize the hyper-parameters by maximizing the probability of drawing these samples with respect to the value of the hyper-parameters:
\begin{equation}
\omega_{\rm opt} = \underset{\omega}{\operatorname{arg\,max}}\left(\log[P_\omega(Y)\cdot P(\omega)]\right) = \underset{\omega}{\operatorname{arg\,max}}\left(-\frac{1}{2}\log(|\mathbf{\Sigma|})-\frac{1}{2}(\mathbf{Y}-\pmb{\mu})^T\mathbf{\Sigma}^{-1}(\mathbf{Y}-\pmb{\mu}) + \log[P(\omega)]\right).
\end{equation}
Here, $P(\omega)$ is the prior distribution of the hyper-parameters (the hyper-prior) and $P_\omega(\mathbf{Y})$ the likelihood of observing $\mathbf{Y}$ given $\omega$. The optimization of the hyper-parameters is the most time consuming computation of the BO process. Therefore, we perform it only after 10, 30, 70, 150, etc. observations of the objective function or if the inversion of $\mathbf{\Sigma}$ fails, which can be an indicator that one of the length scales of the kernel function is too large.

\begin{figure}[htp]
\centering
\includegraphics[width=0.45\linewidth]{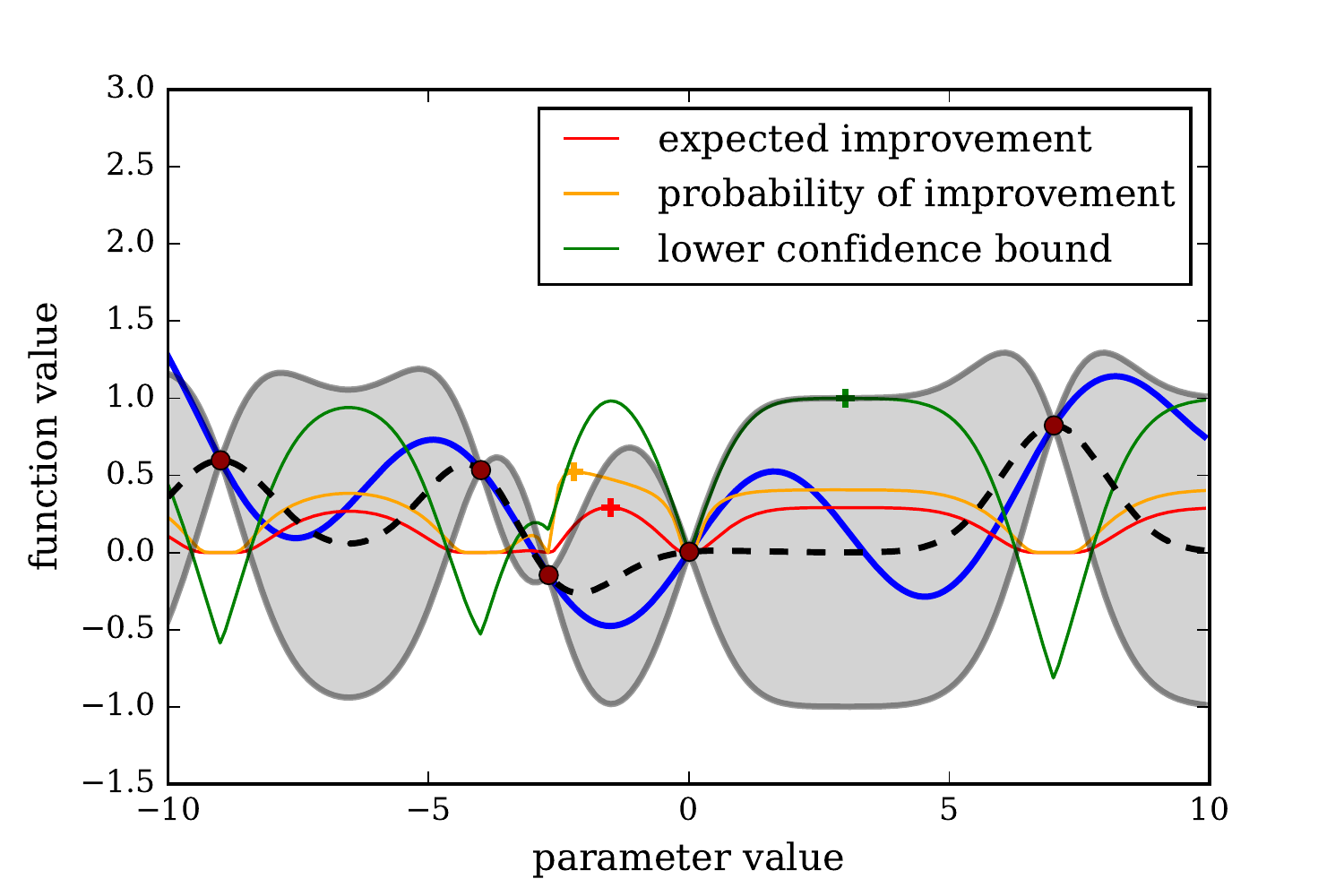}
\includegraphics[width=0.45\linewidth]{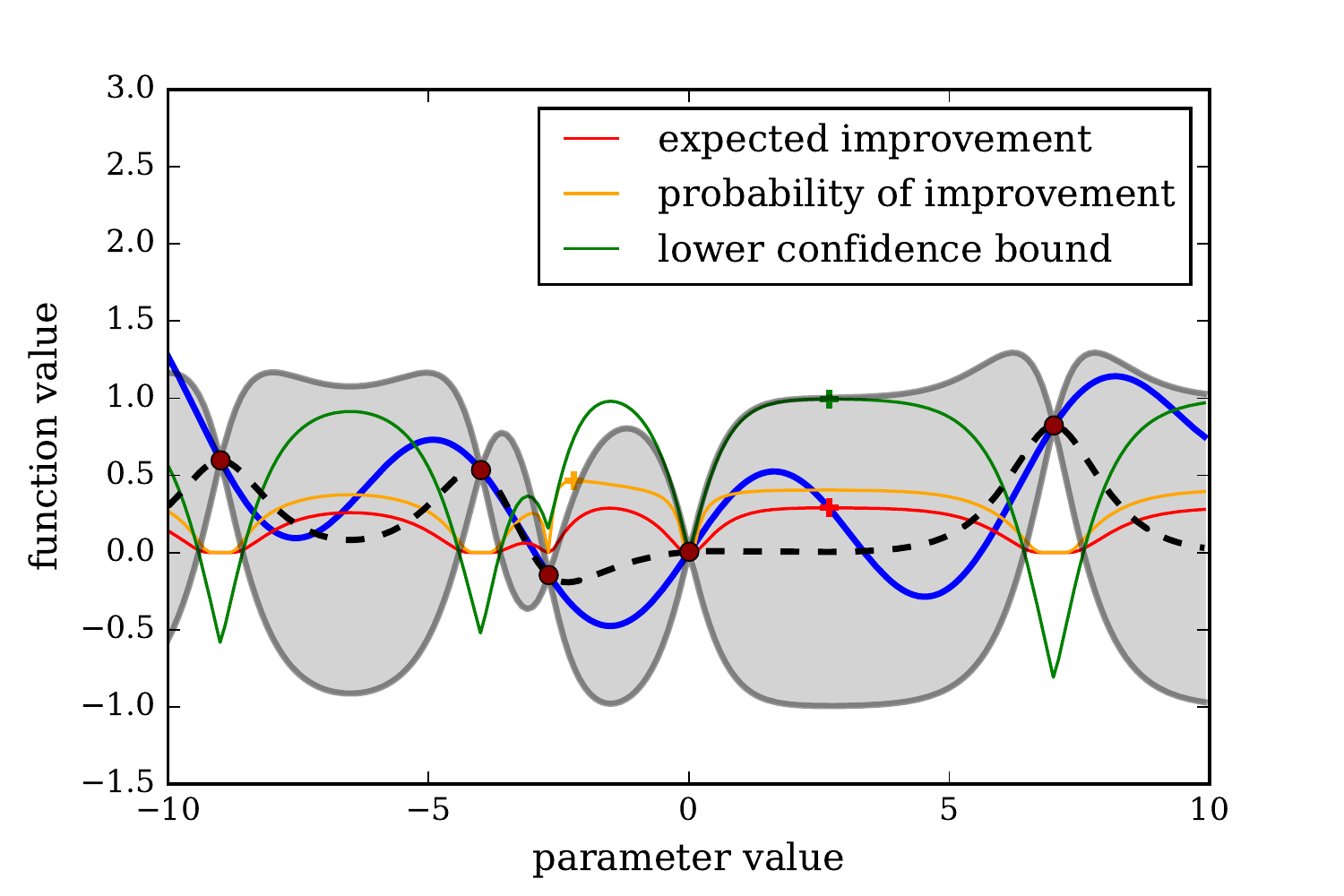}
\caption{The three acquisition functions defined in Eqs.~\eqref{eq:PI}, \eqref{eq:EI}, and \eqref{eq:LB} for the Gaussian processes shown in Fig.~\ref{fig:Regression}. The maxima of the acquisitions functions are marked with crosses.}
\label{fig:Aquisitions}
\end{figure}

\subsection{Acquisition functions}

The acquisition function $\alpha: \mathcal{X} \rightarrow \mathbb{R}$ determines the heuristic search strategy for the minimum of the objective function. Having acquired $N$ data samples at $\mathbf{x}_1,\cdots,\mathbf{x}_N$ with $y_{\rm min} = \min\{f(\mathbf{x_1}),\cdots,f(\mathbf{x_N})\}$ the next calculation is performed at 
\begin{equation}
\mathbf{x}_{N+1} = \underset{\mathbf{x} \in \mathcal{X}}{\operatorname{arg\,max}}\; \alpha(\mathbf{x},y_{\rm min}).
\end{equation}

Many aquisition functions have been proposed so far~\cite{shahriari2016taking}. In the following, three common and rather simple aquisition functions are considered that depend on the local mean value of the GP $\mu_N(\mathbf{x})$ and its local standard deviation $\sigma_N(\mathbf{x})$. Both quantities can be directly derived from the normal distribution of  Eq.~\eqref{Eq:regression} by setting $\mathbf{Y}^* = (G'_{\mathbf x})$.

The considered acquisition functions are the probability of improvement
\begin{equation}
\label{eq:PI}
\alpha_{\rm PoI}(x,y_{\rm min}) = P(f(x) < y_{\rm min}) = 
\frac{1}{2} \left[
    1+ \rm{erf}\left(
       \frac
       	{y_{\rm min} - \mu_N(\mathbf{x})}
       	{\sqrt{2}\sigma_N(\mathbf{x})}
    \right)
 \right],
\end{equation} 
the expected improvement
\begin{equation}
\label{eq:EI}
\alpha_{\rm EI}(x,y_{\rm min}) = \mathbb E[ \max(0,y_{\rm min} - f(x)) ] = 
\alpha_{\rm PoI}(x,y_{\rm min})(y_{\rm min} - \mu_N(\mathbf{x})) + 
 \frac{\sigma_N(\mathbf{x})}{\sqrt{2\pi}} \exp\left(
	\frac
       	{(y_{\rm min} - \mu_N(\mathbf{x}))^2}
       	{2\sigma_N(\mathbf{x})^2} 
 \right),
\end{equation} 
and the lower confidence bound
\begin{equation}
\label{eq:LB}
\alpha_{\rm LCB}(x) = - (\mu_N(\mathbf{x}) - \sigma_N(\mathbf{x})).
\end{equation} 

The three acquisition functions are compared in Fig.~\ref{fig:Aquisitions}. Although all functions have local maxima at similar parameter values, the different height of the maxima can lead to substantially different search strategies.

\subsection{Parallel evaluations}

Often parameter scans of objective functions are performed in parallel, e.g., on a computer cluster. The standard approach of BO is to include the knowledge of all $N$ previous function evaluations in order to determine the parameter values of the $N+1$-st evaluation. This, however, hinders the parallelization of BO. In order to start a new computation while $M$ computations with the parameters $\mathbf{X} = \{\mathbf{x}_1,\cdots,\mathbf{x}_M\}$ are still running, one has to incorporate the running computations into the acquisition functions such that parameter values close to those in $\mathbf{X}$ are avoided. The statistical approach to marginalize the acquisition function over all possible outcomes of the running computations is computationally expensive. 
Therefore, we follow a proposal by Gon\-z\'ales~\emph{et~al.}~\cite{pmlr-v51-gonzalez16a} 
where a penalization function $0\leq \phi(\mathbf{x},\mathbf{x}^*)\leq 1$ is introduced that can be used in conjunction with the acquisition functions. 
In their work it was defined for maximizing a objective function. We use a corresponding expression for minimization
\begin{equation}
\phi(\mathbf{x},\mathbf{x}^*) = \frac{1}{2}\rm{erfc}(-z)\;\; \text{with}\;\;
z = \frac{1}{\sqrt{2}\sigma_N(\mathbf{x}^*)}\left( L \| \mathbf{x}-\mathbf{x}^* \| + \mu_N(\mathbf{x}^*) - y_{\rm min} \right),
\end{equation}
where $L$ is the Lipschitz constant of the objective function that can be estimated by $L = \max_{\mathcal X} \|\nabla \mu_N(\mathbf{x})\|$. In order to incorporate different length scales of the parameters, we use the norm $\| \mathbf{x}-\mathbf{x}'\|^2 = r^2(\mathbf{x},\mathbf{x}')$, where $r^2$ was defined in Eq.~\eqref{eq:SE}. In the numerical experiments of Sec.~\ref{sec:experiments} we restrict the number of evaluations of the objective function running in parallel to 12.

\section{Application to a toy example:\\ Meta surface shape optimization}
\label{sec:Surface}

\begin{figure}[b!]
	\centering
		\includegraphics[width=0.6\textwidth]{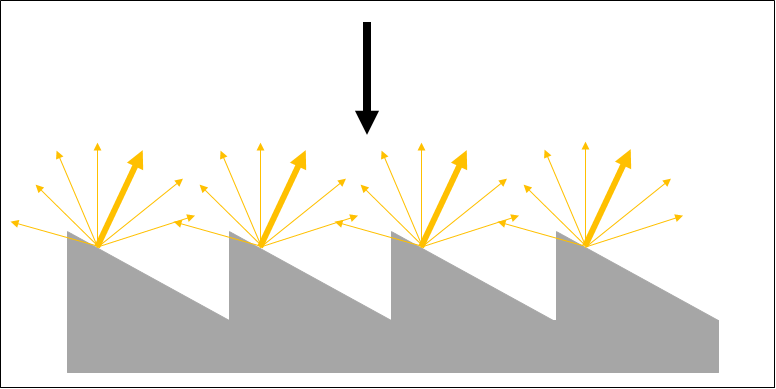}
	\caption{Schematic of the system. Light with a wavelength of $\mathrm{\lambda}$ = 450 nm illuminates the periodic meta-mirror surface under normal incidence, which produces several diffraction orders. The objective is to find the actual shape of the mirror that maximizes the reflectance into the first diffraction order.}
	\label{fig:model_schematic}
\end{figure}

As a toy example, we aim to optimize the shape of a reflective meta surface. It  consist of a 1D-periodic surface structure (refractive index $n=0.4 + 5.0i$) with a periodic length $L_x = 1350$\,nm. The objective is to find a shape of the surface maximizing reflectance into the first diffraction order. Figure~\ref{fig:model_schematic} shows a schematic sketch of the system. The surface is illuminated with $P$-polarized light ($\mathrm{\lambda} = 2\pi/k_0$ = 450 nm) at normal incidence, producing seven reflected diffraction orders corresponding to the diffraction indices $n$ going from -3 to 3,

\begin{equation}
\mathbf{E}_{\rm reflected}(\mathbf{r}) = \sum_{n=-3}^{3}{\mathbf{A}_n e^{i\mathbf{k}_n\cdot \mathbf{r}}}\;\; \text{with}\;\; \mathbf{k}_n = \left(\frac{2\pi n}{L_x},0,\sqrt{k_0^2-\left(\frac{2\pi n}{L_x}\right)^2} \right).
\end{equation}

The objective is to increase the reflectance into the first order ($n = 1$), it can be expressed as
\begin{equation}
O = -|A_1|^2\frac{k_{1,z}}{k_0}.
\end{equation}
Here, the power in the first diffraction order is multiplied by -1 in order to convert the optimization problem into a minimization problem.

In order to allow for arbitrary, smooth and nano-structured surfaces, the surface shape is parameterized using a B-spline expansion.
Using $N_{\rm B}$ B-splines with equidistant knots, the surface elevation is be expressed as 
\begin{equation}\label{eq:surface}
z(x) = \sum_{i=1}^{N_{\rm B}} {A_i B_i(x)}.
\end{equation}

\begin{figure}[htp]
	\centering
		\includegraphics[width=0.35\textwidth]{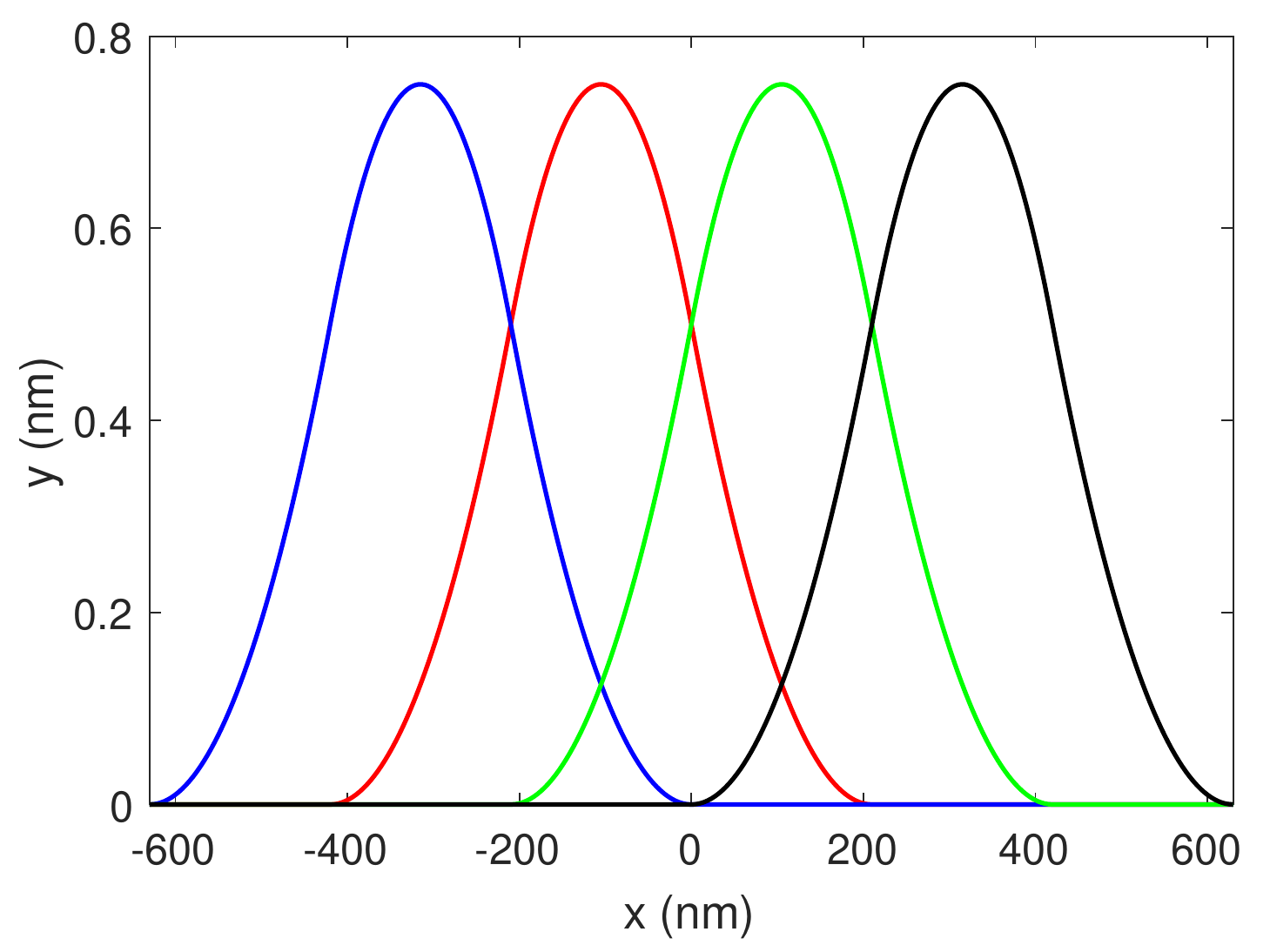}
	        \caption{B-splines used for parameterizing the shape of the surface.
                  The shape will be generated by a linear combination of four (resp.~six) third-order B-splines.}
	\label{fig:B-splines}
\end{figure}

\begin{figure}[htp]
	\centering
		\includegraphics[width=0.7\textwidth]{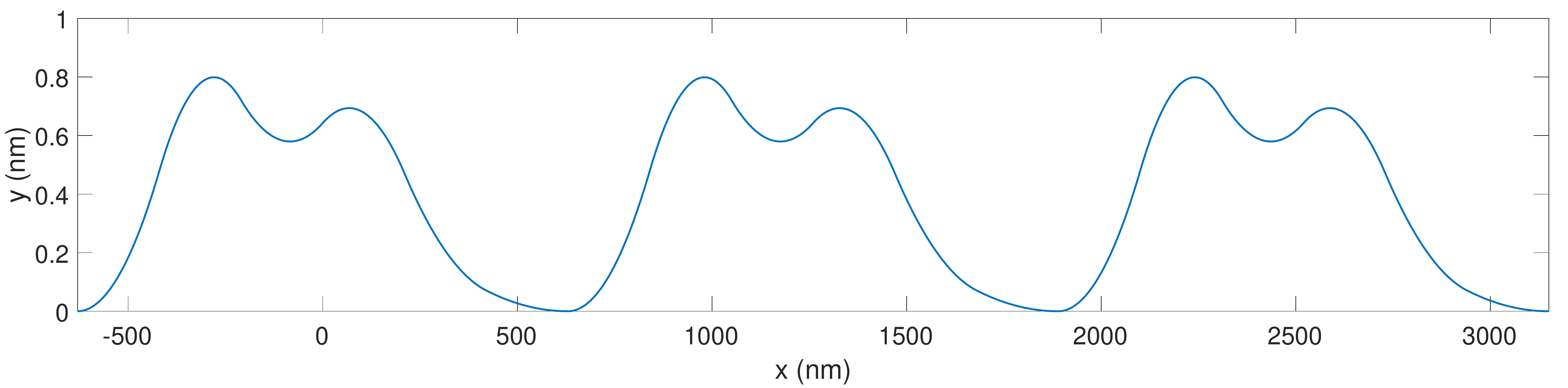}
		\includegraphics[width=0.7\textwidth]{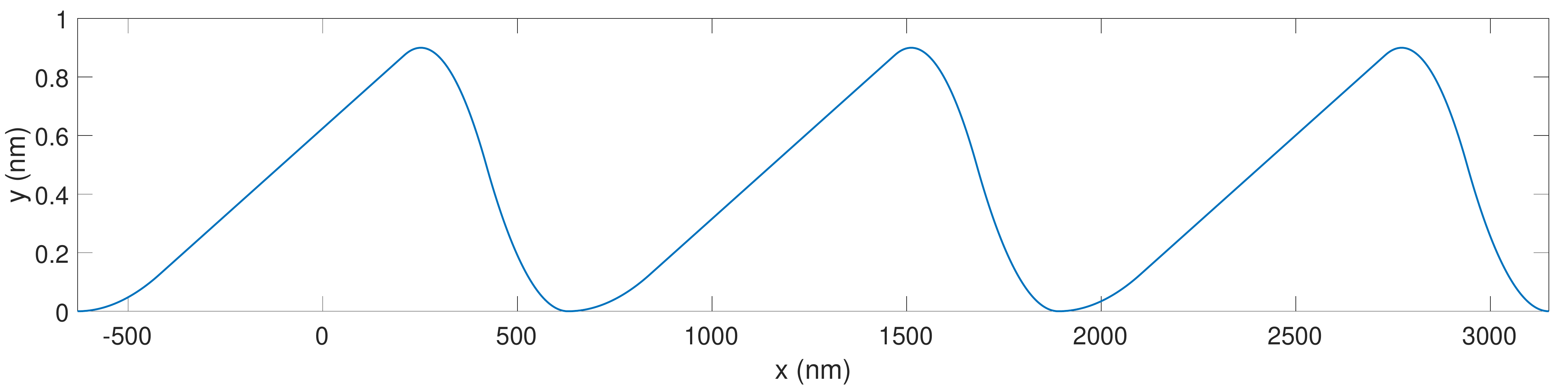}
	\caption{Examples of shapes obtained as linear combinations of the B-splines shown in Figure~\ref{fig:B-splines}.}
	\label{fig:examples_b_spline}
\end{figure}

Figure~\ref{fig:B-splines} shows each of the B-splines used for the case $N_{\rm B} = 4$.
The use of B-splines ensures continuity of the surface. Moreover, it allows for a simple surface description and is flexible in the sense that many different shapes can be described with few parameters.
Figure~\ref{fig:examples_b_spline} shows two different surface shapes (three periods of the periodic surface)
obtained using Eq.~\eqref{eq:surface} with $N_{\rm B} = 4$ and for different values of the amplitudes $A_i$.
Clearly, the chosen basis does not incorporate the physical intuition  that a simple structure of flat inclined surfaces
(as shown schematically in Figure~\ref{fig:model_schematic}) allows
also for high reflectance into a specific diffraction order (for our setup $O=0.85$).
Moreover, the use the B-Spline basis leads to the existence of many local minima, which can make it difficult to find the global minimum. 
However, since the aim of this contribution is to evaluate BO in the context of a global optimization problem,
we do not constrain the parametrization of the surface to match the physical intuition.
This allows for meaningful comparison of various BO approaches. 

\section{Numerical experiments}
\label{sec:experiments}

\subsection{Comparison of BO approaches}

The aim of the numerical experiments presented here is to compare the performance of six different BO approaches. The different approaches are distinguished in that they either use the squared exponential kernel $k_{\rm SE}$ or the Mat\'{e}rn $5/2$ kernel $k_{\rm M 52}$ defined in Eqs.~\eqref{eq:SE} and~\eqref{eq:M52} in conjunction with the three different acquisition functions $\alpha_{\rm PoI}$, $\alpha_{\rm EI}$, and $\alpha_{\rm LCB}$ defined in Eqs.~\eqref{eq:PI},~\eqref{eq:EI}, and~\eqref{eq:LB}. 

The BO procedure is implemented in Python while the light scattering is simulated with the finite-element solver JCMsuite~\cite{Pomplun2007pssb}. 
In order to define the Gaussian process, we used the python module \texttt{gptools}~\cite{chilenski2015improved}.

In many optimization tasks the question arises how many degrees of freedom should be considered. Generally, a higher number of free parameters can lead to better optimization results. However, the search space increases exponentially with more degrees of freedom, such that a restriction to a lower number might be favorable. In order to consider this problem in the context of BO, the example problem is optimized with both $N_{\rm B} = 4$ and $N_{\rm B} = 6$ parameters describing the free-form surface. In order to determine the averaged performance of the different approaches each BO has been repeated for 10 times. 

\begin{figure}[ht]
\centering
\includegraphics[width=0.85\linewidth]{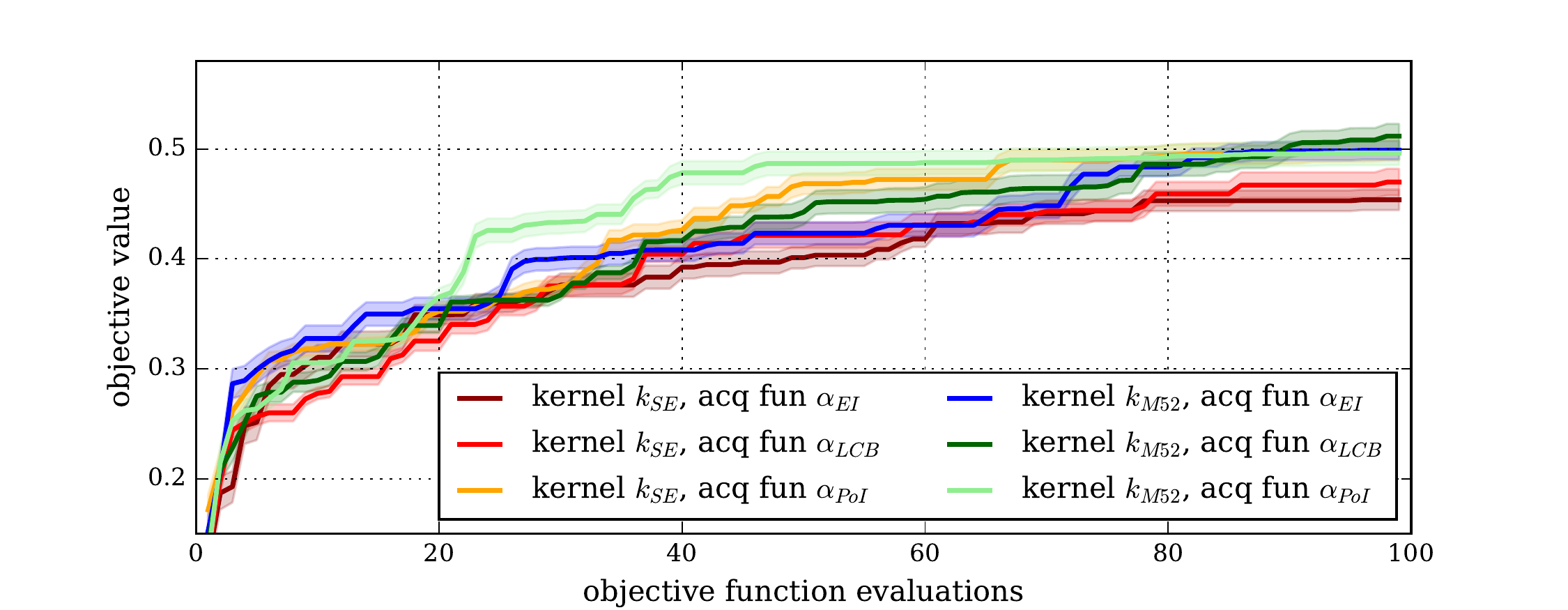}
\includegraphics[width=0.85\linewidth]{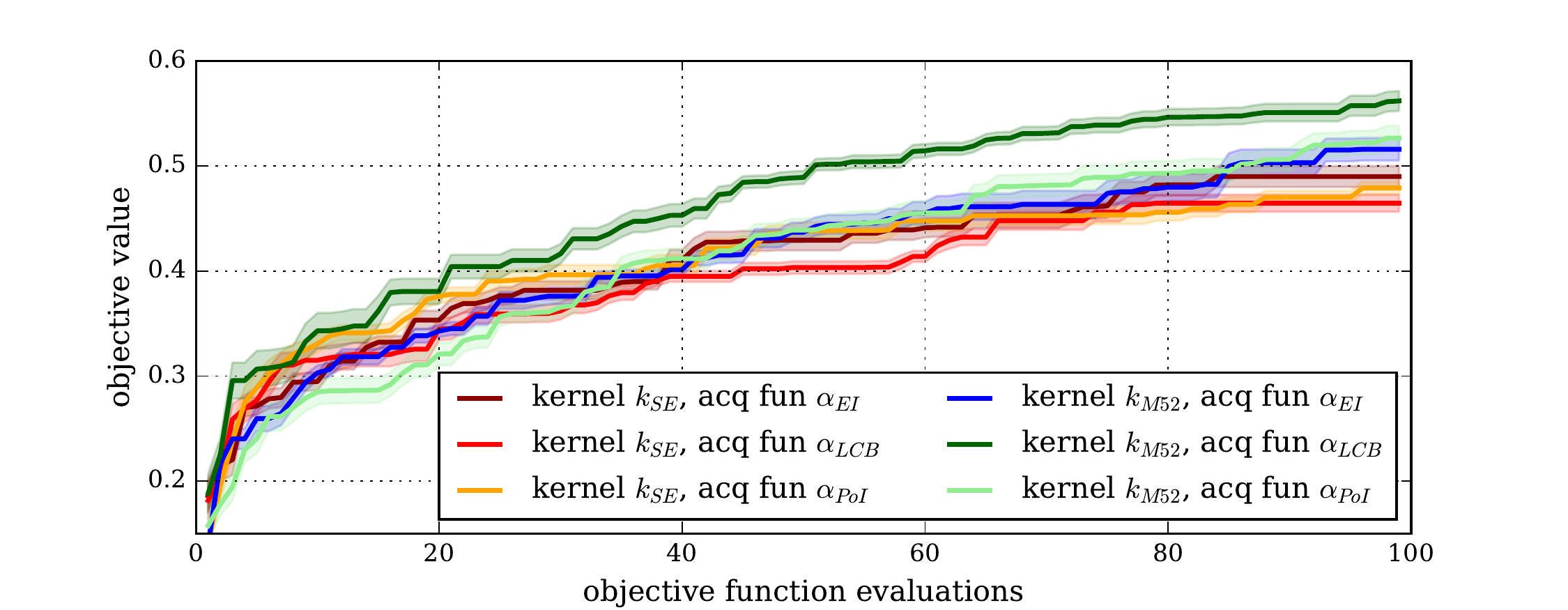}
\caption{Comparison of the averaged performance of different optimization strategies as a function of the number of objective function observations.
The width of the shaded regions corresponds to the standard error. \textbf{Top}: Optimization of the objective function with $N_{\rm B} = 4$. 
\textbf{Bottom}: Optimization of the objective function with $N_{\rm B} = 6$.}
\label{fig:Optimization}
\end{figure}

Figure~\ref{fig:Optimization} shows the performance of each approach as a function of the number of objective evaluations. The results of the optimization after acquiring 100 samples of the objective function are shown in Tab.~\ref{tab:comparison}. The Mat\'{e}rn $5/2$ kernel $k_{\rm M 52}$ leads to slightly  better optimization results. However, within the standard deviation all approaches are performing quite similarly.

Although the search space is much larger when optimizing 6 free parameters, the optimization results are similar. Therefore, it is favorable to include more parameters in the optimization since on the long run (many objective evaluations) the optimization with 6 free parameters will of course eventually outperform an optimization with 4 free parameters.

\begin{table}[ht]
\begin{center}
\begin{tabular}{|p{12em}| l l l| l l l| }
\hline 
Kernel & \multicolumn{3}{ |c| }{$k_{\rm SE}$} & \multicolumn{3}{ c| }{ $k_{\rm M52}$}\\
Aquisition function & $\alpha_{\rm EI}$ & $\alpha_{\rm LCB}$ & $\alpha_{\rm PoI}$ & $\alpha_{\rm EI}$ & $\alpha_{\rm LCB}$ & $\alpha_{\rm PoI}$ \\
\hline 
\hline 
4 parameters & $0.45\pm0.06$ & $0.47\pm0.08$ & $0.50\pm0.06$ & $0.50\pm0.06$ & $0.51\pm0.07$ & $0.50\pm0.07$ \\
6 parameters & $0.49\pm0.06$ & $0.46\pm0.05$ & $0.48\pm0.04$ & $0.52\pm0.07$ & $0.56\pm0.06$ & $0.53\pm0.07$ \\
\hline 
\end{tabular}
\vspace{1em}\caption{Average optimization results for different kernels and acquisition functions. For each algorithm the mean and standard deviation of the best result found after 100 iterations is shown.}
\label{tab:comparison}
\end{center}
\end{table}

\subsection{Optimized surface structure}

\begin{figure}[ht]
\centering
\includegraphics[width=0.6\linewidth]{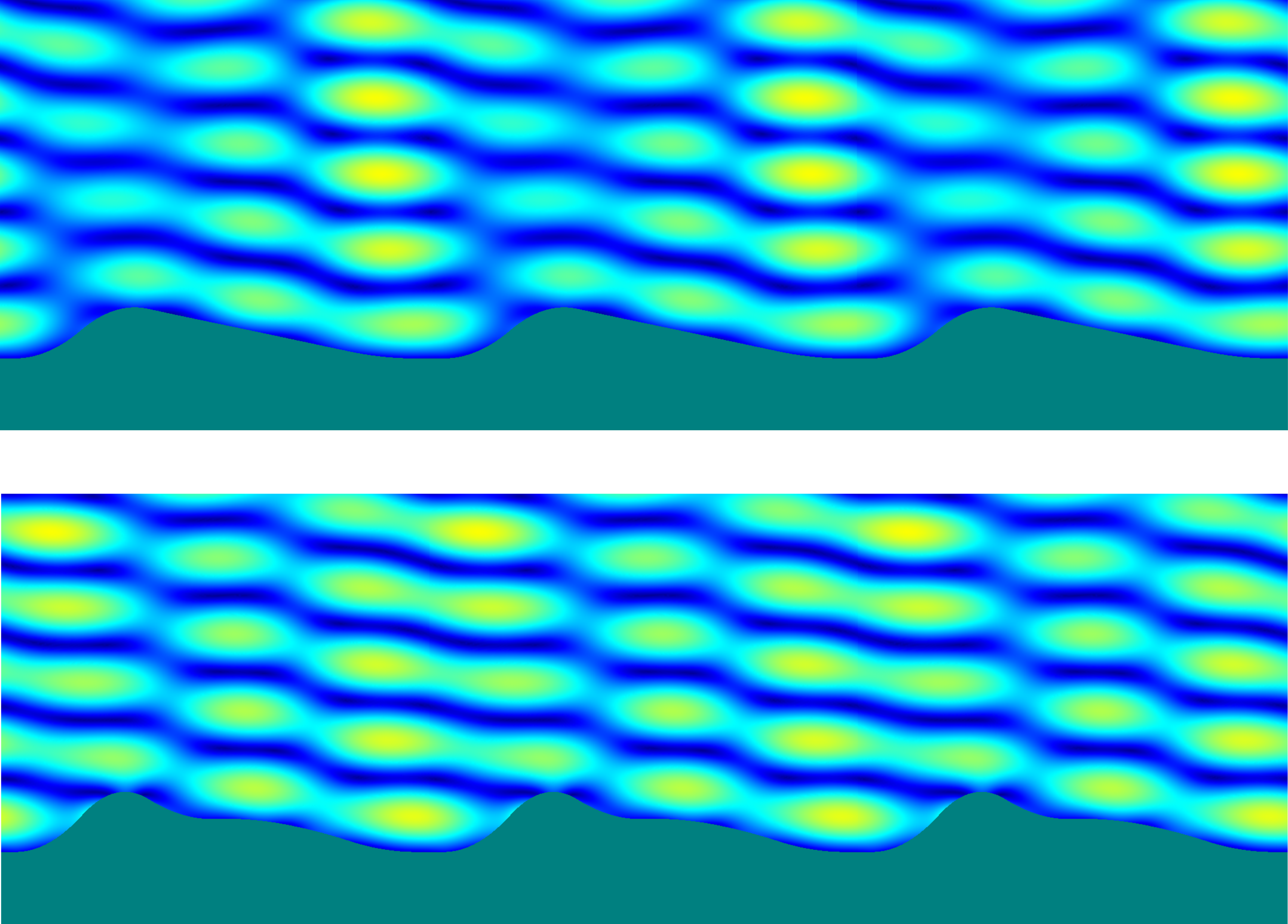}
\vspace{1em}
\caption{Density plot of the $x$-component of the electric field $|\mathbf{E_x}(x,y)|$ above the meta-surface (turquoise). \textbf{Top:} Optimal result for a flat inclined surface created with $N_{\rm B} = 4$ B-splines. \textbf{Bottom:}  Optimal BO-result with $N_{\rm B} = 4$ free parameters. The results of the objective function are 0.59 (top) and 0.67 (bottom).}
\label{fig:Fields}
\end{figure}

Within our framework, a physically intuitive approach to create an optimized structure is to consider a flat surface with a certain inclination angle. Using $N_{\rm B} = 4$ B-splines to create this surface, we obtain an optimal result of $O = 0.59$. For the Bayesian optimization we did purposely not include this physical intuition.
However, the best result obtained with BO and $N_{\rm B} = 4$ free parameters reaches a reflectance into the first order of $O=0.67$, i.e., a higher value. 
Figure~\ref{fig:Fields} shows the corresponding surfaces and field distributions.

\section{Conclusion}
\label{sec:conclusion}

Bayesian optimization allows for a systematic optimization of systems with many degrees of freedom. Although the performance varies over different approaches, all of them reach a comparable level of optimization such that there is no need for \emph{a priori} knowledge of the best BO approach.
In the future we plan to extend the approach to include the observation of derivatives of the objective function with respect to the design parameters.
This is promising since, on the one hand, it is possible within the finite-element framework for solving Maxwell's equations to compute partial derivatives
rigorously without significant additional computational costs~\cite{doi:10.1117/12.2011154}. 
On the other hand, Gaussian processes can be extended to include also derivative information~\cite{murray2003derivative}. 
This additional information might allow for search strategies that reach optimal results with significantly fewer evaluations of the objective function.

\section*{Acknowledgments}
This project has received funding from the Senate of Berlin (IBB, ProFIT
grant agreement No 10160385, FI-SEQUR), co-financed by the European Fund 
for Regional Development (EFRE).
This project has received funding from the European Union’s Horizon 2020 
research and innovation programme under the
Marie Sk{\l}odowska-Curie grant agreement No 675745 (MSCA-ITN-EID NOLOSS).

\bibliography{bibl}
\bibliographystyle{spiebib}

\end{document}